# Topology of Roscoe's- and Hvorslev's- Surfaces in the Phase Space of Soil Mechanics :

## P. Evesque

Lab MSSMat, UMR 8579 CNRS, Ecole Centrale Paris
92295 CHATENAY-MALABRY, France, e-mail evesque@mssmat.ecp.fr

**Abstract:**
*In general, the evolution of soil submitted to simple stress-strain paths is characterized using the 3d phase space* (v,p',q) *i.e. (specific volume, mean intergranular pressure, deviatoric stress); one finds that all trajectories end up at a line of attracting point called the critical-state line. The surface of Roscoe (Hvorslev) is defined as the surface made of the last part of the set of trajectories ending to a given critical point and coming from all states of normally consolidated soils (from all states of over consolidated soils). It is demonstrated that these two sets are part of the same surface, which is confirmed by experimental data.*

______________________________________________________________________

One of the most common apparatus used to characterise the mechanical behaviour of granular materials or clays is the classical axis-symmetric triaxial set-up, which allows to apply different loading ($u_w$, $\sigma_1$, $\sigma_2=\sigma_3$) to samples and to measure the volumetric ($\epsilon_v$) and axial ($\epsilon_1$) strains, the evolution of the specific volume v; here $u_w$ stands for the water pressure and $\sigma_1$, $\sigma_2=\sigma_3$ are the total axial and radial stresses. Coherent results are obtained when inter-granular stresses defined as $\sigma'_i=\sigma_i-u_w$ are used; it is even better to work with the inter-granular mean and deviatoric stresses defined respectively by $p'=(\sigma'_1+\sigma'_2+\sigma'_3)/3$ and $q=\sigma'_1-\sigma'_2$ and to plot q/p' and v (or $\epsilon_v$) as functions of $\epsilon_1$ in order to characterise the behaviour of the sample. This is recalled in Fig. 1. It is worth noting that under simple stress paths the mechanical behaviour of the samples tends to an asymptotic behaviour at large deformation which is independent of initial conditions ($v_o,p'_o,q_o$) and is characterised by a pure friction behaviour (q/p' =M= constant independent of p' , v independent of p' but depending of p'). This asymptotic state is known as the "critical" state [1-4] and is characterised by ($v_c(p'_c),p'_c,q_c=Mp'_c$). However, as this state has nothing to do with the critical point of a phase transition nor with a critical bifurcation I will quote it with inverted comma. It can be seen just as an attracting point (or attractor) in the language of dynamical system theory [5-9].

The way a granular system or a clay reaches the "critical" state under a continuous axial loading is twofold depending on whether it contracts always or if it needs to dilate during some part of its evolution. For clay, these two cases are labelled normally- or over- consolidated. For sand and granular material, I will called them loose and dense cases.

Since Hvorslev and Roscoe at least, one has used also the 3d space (v,p',q) to characterise the evolution of a sample. This number of coordinates seems to be sufficient to describe the evolution, so that it can be considered as the phase space (in the thermodynamics terminology). In this space, the "critical" state is the intersection between two surfaces ($v_c(p')$ and $q_c=Mp'$); it is then a curve of dimension 1; this curve is approximately a straight line in the system $(v,\ln(p'), \ln(q))$ [10]. However, projections on planes $(v,\ln(p'))$ and $(q,p')$ are used in general.

So, Hvorslev and Roscoe have tried and understood how the material reaches the critical state, Roscoe in the case of loose or normally consolidated materials, Hvorslev for dense or over consolidated ones. They have both concluded that the ensemble of trajectories leading to a given "critical" state pertained to a surface called the surface of Hvorslev (Roscoe) for dense (loose) materials. They are generally represented as distinct surfaces, making a given angle between them (see Figures 11.13-11.16, 12.8 & 13.4 of ref. [1] for instance).

The aim of this paper is to show that these two surfaces pertain to a unique surface. Following previous works [5-7, 9], we will use for this the language of dynamical-system theory [8] for which the $(v,\ln(p'),q)$ space is the phase space; it is of dimension $d_e=3$. The ensemble of "critical" states is an attracting curve of dimension $d_a=1$, which attracts its neighbourhood. The evolution of the system is characterised by a curve starting from an initial point $(v_o,\ln(p'_o),q_o)$ and ending at $(v_c,\ln(p'_c),q_c)$. In this scheme, each trajectory is parametrized by the axial deformation $\varepsilon_1$, which plays the role of time in classical dynamical-system theory hence.

*Theoretical argument:*

We have seen that the "critical" state line is an attractive line; but such a line cannot attract its neighbourhood parallel to itself, otherwise it will end up as an attracting point. So, this imposes that the directions of attraction define a space whose dimension $d_b$ be smaller or equal to $d_e-d_a$. The equality shall be observed if the attractive line attracts its whole neighbourhood, which is the case here. So, $d_b=d_e-d_a=2$; it is then a field surface which cuts the attractor line at the considered attractor point. Furthermore, two field surfaces arriving at two different attractors closed to each other shall not cross in order to maintain single evolution; this implies that field surfaces are parallel and that their normal vector at the "critical" point shall vary continuously. The field surface corresponding to a "critical" state shall be identified to the Roscoe's (and to the Hvorslev's) surfaces for the loose (and dense states respectively).

*Few remarks and discussions:*
1) This is a "natural" result : the phase space is 3d, the surfaces of Hvorslev and Roscoe are 2d and have a common point (the "critical" state), they shall have a common line which can be a border line, so that these two surfaces define a single surface anyhow.
2) However, this border line could be the location of an angular discontinuity , indicating a discontinuity of the vector normal to the surface at this location.
3) There could be an other kind of discontinuity: each point of the attractor line could be the location of an angular discontinuity for each field surface, if each field surface looks topologically like a cone whose summit is the attractor point.

4) In general, in the theory of dynamical systems, such defects result from a projection of the complete phase space on a sub-space of less dimensionality [8]; then it should indicate that the phase space used here (v,ln(p'), q) is not complete and that the real phase space is larger. This is why it is quite important to study the manner the trajectories arrive at the "critical" state in the (v,ln(p'),q) space.

In general, in the theory of dynamical systems, such defects result from a projection of the complete phase space on a sub-space of less dimensionality [8]; then it should indicate that the phase space used here (v,ln(p'), q) is not complete and that the real phase space is larger. This is why it is quite important to study the manner the trajectories arrive at the "critical" state in the (v,ln(p'),q) space.This is what we will do now on; hopefully there are a series of results one can find in the literature; we will use those given by Biarez & Hicher [4] and by Saïm [11]; in particular this last author has collected a series of results from tests performed by different teams (Al Issa, Al Mahmoud & Sharour, Bousquet & Flavigny, Bouvard, Charif, Colliat, Desrues & Mokni, Djeddid & Flavigny, Khazar & Flavigny, Ladd, Lee & Seed, Lefevbre & Flavigny, Meghachou & Flavigny, Mokham(1983) , Naskos, Perez, , Trueba, Zervoyannis, Ziani).

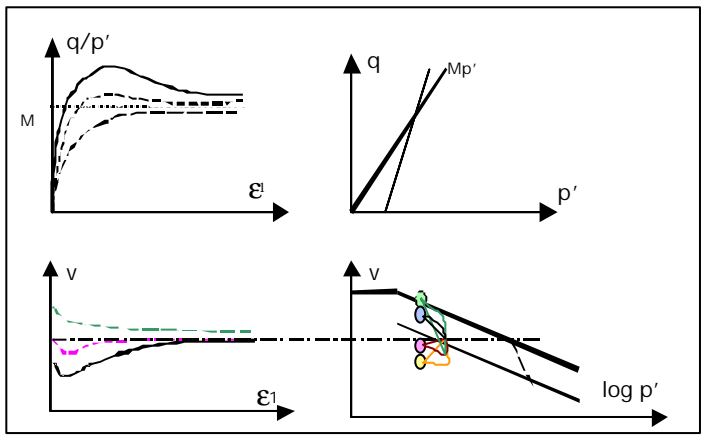

**Figure 1:** Typical triaxial test results: $\sigma'_2=\sigma'_3$=constant triaxial tests at the same initial pressure $p'_o$ for three initial densities.

## *Comparison with experimental data:*

In order to demonstrate the equivalence between the two surfaces, one is led to demonstrate that the trajectories arriving at the "critical" state arrives just in opposite direction for the two surfaces, i.e. for dense and loose packings. Furthermore, as the surface is 2d, one needs only to study this arrival for two independent trajectories and hence for two independent types of tests. We have chosen to use undrained paths (v=constant) and classical drained triaxial paths ($\sigma'_2=\sigma'_3$=constant).

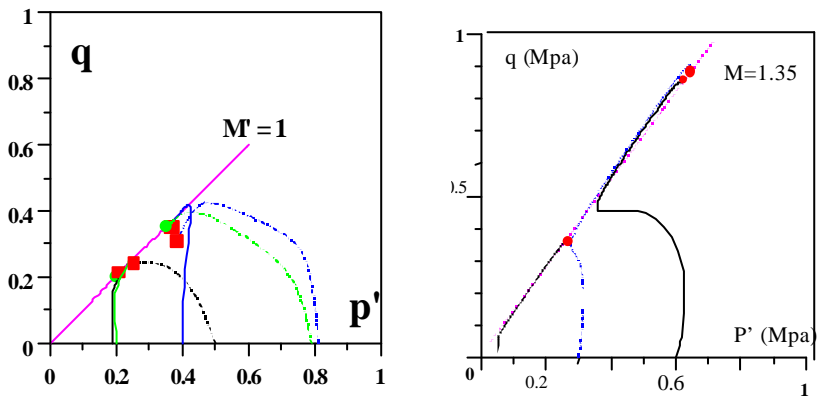

**Figure 2:** Typical triaxial test results: a) and b) $v=c^{ste}$ test for different initial pressure $p'_o$ and specific volumes $v_o$. a) clay; b) Flavigny experiment on Hostun sand.

*Triaxial path* ($\sigma'_2=\sigma'_3=c^{ste}$)*:*

This test imposes $p'=q/3+p'_o$; so the projection of the trajectory on to the (p',q) plane is a straight line imposed by experimental conditions. On this plane, the trajectories for dense and loose sands arrive at the "critical" state just in opposite direction since this arrival occurs by increasing (decreasing) p' and q for loose (dense) sand. Thus, one is faced to demonstrate that the projections of the trajectories arrive in opposite direction in the (v,p') plane too in order to demonstrate that trajectories of dense and loose materials arrive really in opposite directions at the "critical" point. This is just what show Fig. 2 of this paper and Fig. 5.19a&b, 5.25 of [4] for sand, figure 5.24b of [4] for aggregates, figures. 5.5, 5.6, 5.7, 5.16, 5.25, 5.26, 5.39, 5.43 of [11] for sand; comparison of figures 4.9, 5.2 & 5.4 of [11] to figures 3.12 à 3.14 of [11] confirm the analysis for clays too. As a matter of fact, agreement looks better when using (v,ln(p')) coordinates; this is quite normal since the "critical" state curve is a straight line with this system of coordinates. Endly, it is worth noting that experimental data exhibit a large uncertainty on the "critical" specific volume; this is linked most likely to the difficulty to generate homogeneous large deformation and hence to approach the "critical" state line. Under these circumstances, it is worth emphasising the quality of the experimental data obtained by Bouvard (Fig. 5.6 & 5.7 of [11]).

$v=c^{ste}$ *(undrained) path:*

As $v=c^{ste}$, the projections of the trajectories on the (v,p') and (v,q') planes are two segments of horizontal lines which reach the "critical" state from the right (left) for loose (dense) materials; it remains to study the trajectories in the (p',q) plane.
For sand, one finds that all trajectories start vertically (p'=constant) [12], then they are curved towards the left (towards smaller p') to join the q=Mp' line; during this part of the curve, axial deformation $\varepsilon_1$ remains small ($\varepsilon_1 < 10\%$). When the sand is loose, (Figs. 3.25 to 3.27 of [11]), the trajectories goes on towards smaller p' and joins the q=Mp' line tangentially to this line and stops. When the sand is dense, (Figs. 6.5 to 6.7 & 6.10 to 6.24 of [11]), the trajectory reaches the q=Mp' line and follows it in the direction of increasing p'.

In the case of clays, (Fig. 2a of this paper, Figs. 3.21 to 3.23, 6.1 & 6.2 of [11]), the trajectories are similar.

It is worth noting that the trajectories of undrained tests i) start obeying p'=$c^{ste}$ and ii) end following the q=Mp' line can be explained with a very simple model [12].

One shall note however, i) that sometimes the trajectory of over-consolidated clays is vertical on a longer path before reaching the q=Mp' line, so that it can reach the q=Mp' line with a vertical direction, ii) that the arrival at the q=Mp' for normally or slightly over consolidated clays is perpendicular to this line and turn left on this line after a large stay at the corner (see Figs. 3.21 to 3.23 of [11]). At first sight, such an angular point does not look physical, especially as the (q,$\varepsilon_1$) curve is smooth; thus it may correspond to an optical effect linked to a bad projection; in this case this will prove the need of the use of a fourth coordinate to define the complete phase space; this last coordinate could be the anisotropy of the contact distribution. Nevertheless, one cannot omit other possibilities which explain the existence of this angular point.

Let us now consider that the trajectory stops when arriving at the q=Mp' line for normally consolidated clays, and then turn left later on this q=Mp' line when sample deformation is proceeded further; indeed this is observed experimentally, but it is not clearly admitted as one can learn from text book; so, this is an important hypothesis which shall be tested in the future, even if it is quite difficult since this part of the curve is unstable.

Nevertheless, this singular behaviour is in agreement with the previous analysis using the (v,ln(p'),q) phase space since it predicts that the trajectory turn right or left on the q=Mp' line depending on the value of the initial specific volume $v_o(p'_o)$ compared to a given value $v_1(p'_o)$ : turn left if $v_o(p'_o) < v_1(p'_o)$, turn right if $v_o(p'_o) > v_1(p'_o)$. Furthermore, as this initial value imposes the left or right turn at the turning point, one may predict that there is a range of initial condition around the critical value $v_1(p')$ for which the result may be quite sensitive to the precise material history and to initial heterogeneities; this lets predict fluctuations of trajectories from test to test; conversely, analysis of these fluctuations may help understanding how heterogeneities interact. Similar remark could have been done for the experiment at $\sigma_1=\sigma_2=$constant, but results look less sensitive, even if one observes in this case that trajectories starting quite near from the "critical" state line in the (v,ln(p')) plane is not captured at once by the attractor line but follows a trajectory parallel to it during some time before turning up suddenly to rejoin it; this might be the expected effect.

So, one may say that trajectories corresponding to undrained conditions end at the "critical" state with a direction given by the set of two equations q=Mp' and v=constant, whatever the initial value of $p'_o$ and $v_o$; these trajectories arrive at this point in opposite direction i.e. towards increasing (decreasing) p' for dense (loose) materials. However, one shall also indicate few slight experimental contradictions: in Fig. 3.21-3.23 of [11], the slope at the end of trajectories are not strictly parallel to the q=Mp' line; this is linked probably to some mechanical instability; in the same way, in Figs. 6.5 to 6.24 of [11], one can see at the very end of the trajectories a diminution of q and p' in the case of quite dense samples; this is also the signature of an other mechanical instability generated by hetereogeneities induced by large deformations.

*Discussion and conclusion:*

To the best of my knowledge, the equivalence between these two surfaces (of Roscoe and Hvorslev) has never been claimed by the experimenters, so that they could not use this point to determine the "quality" of their results . So observing these facts demonstrate the quality of the experiments together with their coherence and prove the relevance of the soil mechanics approach.

Furthermore, it is worth noting that this relevance is observed also for systems which are unstable, and which are maintained in this unstable situation owing to the technique of the experimenter who have prepared the sample "correctly". This last remark could be an important theoretical objection, since one could argue that the unstable stable should not be observed. It is not however, since other examples are known in physics and in mechanics where one knows how to keep a system in an unstable state and to perform experiment on it: i) one knows how to prepare cooled water (T<0°C) in an overmelted state; this is an unstable state since it can freeze at any time if one perturbs it in a wrong way; ii) buckling of tubes under axial pressure can be delayed above the threshold pressure $p_t$ , but this state is unstable....

The 3d phase space (v, ln(p'), q) seems to be large enough to describe the evolution of the mechanical behaviour of granular materials or clays under simple stress paths in most cases. This looks quite surprising since one knows that a granular material does not remain isotropic as the deformation proceeds, even when it remains homogeneous: contact distribution evolves and becomes anisotropic. However, if the distribution depends essentially on the deformation $\varepsilon_l$, or on the q/p' ratio the description in terms of (v,ln(p'),q) can remain sufficient if one takes into account (i) the experimental uncertainty, (ii) the fact that the "critical" state has a well defined constant anisotropy which depends on the direction of the major principal stress only and (iii) that one does not try to demonstrate that two different trajectories which converge to the same "critical" state never cross each other whatever its initial anisotropy or (iv) that one does not want to study singular behaviours such as that one observed in undrained experiment on normally consolidated clays. Indeed, points (iii) and (iv) would require to introduce the contact anisotropy as a new parameter and as a new dimension. In the same way, one shall introduce other new coordinates in order to make the description more complete and usable in non axi-symmetric test. This would render the system more intricate, making it difficult to use by experimenters, without improving it so much probably.

*Conclusion* : In this paper, we propose a new theoretical approach which is based on general consideration of the structure of trajectories in the phase space. It allows to enlighten the structure of the Roscoe's and Hvorslev's surfaces and to show that they are parts of the same surface. It seems to agree with experimental data.

However it is worth noting that part of the experimental data concerning undrained tests and $\sigma_2=\sigma_3=c^{ste}$ compression tests where not included as characteristics of the trajectories in text books on soil mechanics. So the experimental validation of this approach remains to be pursued. In the same way, theoretical soil mechanics is able to describe a remarkable set of features. The compatibility of this new approach with the existing theories shall be demonstrated and their differences shall be discussed. This will be done in forthcoming papers.

*Acknowledgements :* I want to thank MM. J. Biarez, F. Darve et E. Flavigny for fruitful discussions.

The electronic arXiv.org version of this paper has been settled during a stay at the Kavli Institute of Theoretical Physics of the University of California at Santa Barbara (KITP-UCSB), in june 2005, supported in part by the National Science Fundation under Grant n° PHY99-07949.